\documentclass[conference]{IEEEtran}
\ifCLASSINFOpdf
\usepackage{ifpdf}
\usepackage{cite}
\ifpdf
  \usepackage[pdftex]{graphicx}
  \DeclareGraphicsExtensions{.pdf}
\else
  \usepackage[dvips]{graphicx}
  \graphicspath{{../Figures/EPS/}}
  \DeclareGraphicsExtensions{.eps}
\fi
\usepackage[cmex10]{amsmath}
\usepackage{amssymb}
\usepackage{wasysym}
\usepackage{mathrsfs}
\usepackage{stmaryrd}
\usepackage{upgreek}
\usepackage{algorithm}
\usepackage{algorithmic}
\usepackage{array}
\newtheorem{defn} {Definition}

\newtheorem{prop}{Proposition}
\newcommand{\xib}{\mathbf{\upxi}}
\usepackage[belowskip=-10pt,aboveskip=0pt]{caption}
\begin{document}
\title{Selecting Two-Bit Bit Flipping Algorithms for Collective Error Correction}
%
\author{\IEEEauthorblockN{Dung Viet Nguyen, Bane Vasi$\acute{\mathrm{c}}$ and Michael W. Marcellin}
\IEEEauthorblockA{Department of Electrical and Computer Engineering\\
University of Arizona, Tucson, Arizona 85721\\
Email: \{nguyendv, vasic, marcellin\}@ece.arizona.edu}}
\maketitle
\begin{abstract}
A class of two-bit bit flipping algorithms for decoding low-density parity-check codes over the binary symmetric channel was proposed in \cite{NMV_11_ISIT}. Initial results showed that decoders which employ a group of these algorithms operating in parallel can offer low error floor decoding for high-speed applications. As the number of two-bit bit flipping algorithms is large, designing such a decoder is not a trivial task. In this paper, we describe a procedure to select collections of algorithms that work well together. This procedure relies on a recursive process which enumerates error configurations that are uncorrectable by a given algorithm. The  error configurations uncorrectable by a given algorithm form its \textit{trapping set profile}. Based on their trapping set profiles, algorithms are selected so that in parallel, they can correct a fixed number of errors with high probability.
\end{abstract}
\section{Introduction}\label{sect_introduction}
With the introduction of high speed applications such as flash memory, fiber and free-space optical communications comes the need for fast and low-complexity error control coding. Message passing algorithms for decoding low-density parity-check (LDPC) codes such as the sum-product algorithm (SPA) offer very attractive error performance, especially for codes with column-weight $d_\mathrm{c}\geq4$. However, the complexity of these algorithms is still high and the decoding speed is limited, mostly due the fact that the operations at variable and check nodes must be carried out for every edge in the Tanner graph. For regular column-weight-three LDPC codes, which allow lower complexity implementation, message passing algorithms (as well as other classes of decoding algorithms) usually suffer from high error floor. This weakness of message passing algorithms in regular column-weight-three LDPC codes justifies the search for alternatives which offer better trade-offs between complexity, decoding speed and error performance.

Among existing decoding algorithms for LDPC codes on the binary symmetric channel (BSC), bit flipping algorithms are the fastest and least complex. The check node operations of these algorithms are modulo-two additions while the variable node operations are simple comparisons. The simplicity of these algorithms also makes them amenable to analysis. Many important and interesting results on the error correction capability of the serial and parallel bit flipping algorithms have been derived (see \cite{NMV_11_ISIT} for a list of references). Unfortunately, their error performance is typically inferior. As a result, bit-flipping-oriented algorithms have been largely considered to be impractical, even after the introduction of some improved versions, such as the one in \cite{miladinovic}.

In \cite{NMV_11_ISIT}, a class of bit flipping algorithms that employ two bits for decoding LDPC codes over the BSC was proposed. Compared to serial and parallel bit flipping, a two-bit bit flipping (TBF) algorithm employs one additional bit at a variable node and one at a check node. The additional bits introduce \textit{memory} in the decoding process, which slows down the decoding when necessary. Initial results showed that decoders which employ a group of these algorithms operating in parallel lower the error floor while maintaining low complexity. However, in \cite{NMV_11_ISIT} we have not given a complete failure analysis of these algorithms, nor have we established the methodology to derive good algorithms and/or a collection of mutually good algorithms.

In this paper, we provide complete failure analysis for TBF algorithms. More importantly, we give a rigorous procedure to select groups of algorithms based on their complementariness in correcting different error patterns. Decoders that employ algorithms selected using this procedure have provably good error performance and, by the nature of bit flipping, high speed. 

As one can expect, a TBF algorithm (like other sub-optimal graph-decoding algorithms) fails on some low-weight error patterns due to the presence of certain small subgraphs in the Tanner graph. In this paper, we characterize a special class of these subgraphs and refer to them with the common term ``trapping sets.'' Our definition of a trapping set for a given algorithm readily gives a sufficient condition for successful decoding. The set of all possible trapping sets of a given decoding algorithm constitutes the algorithm's trapping set profile. A unique property of trapping sets for TBF algorithms is that a trapping set profile may be obtained by a recursive procedure. The diversity among trapping set profiles of different algorithms allows us to select groups of algorithms such that they can collectively correct error patterns that are uncorrectable by individual algorithms.

The rest of the paper is organized as follows. Section \ref{sect_pre} gives the necessary background. Section \ref{sect_mot} gives motivation. In Section \ref{sect_analysis}, we define trapping sets, trapping set profiles and describe the recursive procedure for constructing a trapping set profile. Section \ref{sect_select} discusses the process of selecting algorithms. Numerical results are given in Section \ref{sect_sim}.
\section{Preliminaries}\label{sect_pre}
%
%
Let $\mathcal{C}$ denote an ($n,k$) binary LDPC code. $\mathcal{C}$ is defined by the null space of $H$, an $m\times n$ \textit{parity-check matrix} of $\mathcal{C}$. $H$ is the bi-adjacency matrix of $G$, a Tanner graph representation of $\mathcal{C}$. $G$ is a bipartite graph with two sets of nodes: $n$ variable (bit) nodes $V(G) = \{1, 2,\ldots, n\}$ and $m$ check nodes $C(G) = \{1, 2,\ldots ,m\}$; and a set of edges $E(G)$. A ($d_{\mathrm{v}}, d_{\mathrm{c}}$)-regular LDPC code has a Tanner graph $G$ in which all variable nodes have degree $d_{\mathrm{v}}$ and all check nodes have degree $d_{\mathrm{c}}$. In this paper, we only consider ($d_{\mathrm{v}}, d_{\mathrm{c}}$)-regular LDPC codes. A subgraph of a bipartite graph $G$ is a bipartite graph $U$ such that $V(U)\subset V(G)$, $C(U)\subset C(G)$ and $E(U)\subset E(G)$. $G$ is said to contain $U$. Furthermore, if $Y$ is a graph which is isomorphic to $U$ then $G$ is also said to contain $Y$. In a bipartite graph $G$, the induced subgraph on a set of variable nodes $V_{\mathrm{s}}\subset V(G)$ is a bipartite graph $U$ with $V(U) = V_{\mathrm{s}}$, $C(U) = \{c\in C(G):\exists v\in V_{\mathrm{s}}\;\mathrm{such~that}\; (v,c)\in E(G)\}$ and $E(U) = \{(v,c)\in E(G):v\in V_{\mathrm{s}}\}$.

A vector ${\mathbf{x}} = (x_1, x_2,\ldots, x_n)$ is a codeword if and only if ${\mathbf{x}} H^\mathrm{T} = \mathbf{0}$, where $H^\mathrm{T}$ is the transpose of $H$. Assume the transmission of the all-zero codeword over the BSC. Denote by $\mathbf{y}$ the channel output vector and denote by ${\mathbf{\hat{x}}}^l = (\hat{x}^l_1, \hat{x}^l_2,\ldots, \hat{x}^l_n)$ the decision vector after the $l$th iteration of the iterative algorithm, where $l$ is a positive integer. At the end of the $l$th iteration, a variable node $v$ is said to be \textit{corrupt} if $\hat{x}^l_v = 1$, otherwise it is \textit{correct}. For the sake of convenience, we let $\mathbf{\hat{x}}^0 = \mathbf{y}$. A variable node $v$ with $\hat{x}^0_v = 1$ is \textit{initially corrupt}, otherwise it is \textit{initially correct}. Let $\mathbf{s}^l = (s^l_1,s^l_2,\ldots, s^l_m)$ denote the syndrome vector of the decision vector after the $l$th iteration, i.e., $\mathbf{s}^l = {\mathbf{\hat{x}}}^l H^\mathrm{T}$. A check node $c$ is said to be satisfied at the beginning of the $l$th iteration if $s_c^{l-1} = 0$, otherwise it is unsatisfied. TBF algorithms are defined as follows.
%
%
\begin{defn}
The class $\mathscr{F}$ of TBF algorithms is given in Algorithm \ref{algo_tbfa}, where $\mathbf{z}^l=(z_1^l,z_2^l,\ldots,z_m^l)$ gives the states of the check nodes at the beginning of the $l$th iteration while $\mathbf{w}^l=(w_1^l,w_2^l,\ldots,w_n^l)$ gives the states of the variable nodes at the end of the $l$th iteration. A variable node $v$ takes its state from the set $\mathcal{A}_{\mathrm{v}} = \{0_{\mathrm{s}},0_{\mathrm{w}},1_{\mathrm{w}},1_{\mathrm{s}}\}$, i.e., it can be strong zero, weak zero, weak one or strong one. A check node takes its state from the set $\mathcal{A}_{\mathrm{c}}= \{0_{\mathrm{p}},0_{\mathrm{n}},1_{\mathrm{p}},1_{\mathrm{n}}\}$, i.e., it can be previously satisfied, newly satisfied, previously unsatisfied or newly unsatisfied. The state $w^0_v$ of a variable node $v$ is initialized to $\Delta_\mathrm{v}(0)\in\{0_{\mathrm{s}},0_{\mathrm{w}}\}$ if $y_v=0$ and to $\Delta_\mathrm{v}(1)\in\{1_{\mathrm{s}},1_{\mathrm{w}}\}$ if $y_v=1$. The state $z^1_c$ of a check node $c$  is initialized to $\Delta_\mathrm{c}(0)\in\{0_{\mathrm{p}},0_{\mathrm{n}}\}$ if $s^0_c=0$ and to $\Delta_\mathrm{c}(1)\in\{1_{\mathrm{p}},1_{\mathrm{n}}\}$ otherwise. A TBF algorithm $\mathcal{F} = (f,l^\mathrm{m}_\mathcal{F},\Delta_\mathrm{v},\Delta_\mathrm{c})$ iteratively updates $\mathbf{z}^l$ and $\mathbf{w}^l$ until all check nodes are satisfied or until a maximum number of iteration $l^\mathrm{m}_\mathcal{F}$ is reached. The check node update function $\Phi:\{0,1\}^2\rightarrow \mathcal{A}_\mathrm{c}$ is defined as follows: $\Phi(0,0) = 0_{\mathrm{p}}, \Phi(0,1) = 1_{\mathrm{n}}, \Phi(1,0) = 0_{\mathrm{n}}$ and $\Phi(1,1) = 1_{\mathrm{p}}$. The variable node update is specified by a function $f:\mathcal{A}_\mathrm{v}\times \Xi_{d_\mathrm{v}}\rightarrow \mathcal{A}_\mathrm{v}$, where $\Xi_{d_\mathrm{v}}$ is the set of all ordered 4-tuples $\xib = (\xi_1,\xi_2,\xi_3,\xi_4)$ such that $\xi_i\in \mathbb{N}$ and $\sum_i{\xi_i}=d_\mathrm{v}$. $\chi_{0_{\mathrm{p}}}^{l}(v), \chi_{0_{\mathrm{n}}}^{l}(v), \chi_{1_{\mathrm{p}}}^{l}(v)$ and $\chi_{1_{\mathrm{n}}}^{l}(v)$ give the number of check nodes with states $z^l_c = 0_{\mathrm{p}}, 0_{\mathrm{n}}, 1_{\mathrm{p}}$ and $1_{\mathrm{n}}$, respectively,  that are connected to $v$. The function $f$ must be symmetric with respect to $0$ and $1$ and must allow every state of a variable node to be reachable from any other state.
\begin{algorithm}
\caption{TBF Algorithm}\label{algo_tbfa}
\begin{algorithmic}\label{algo_tbfa2}
\STATE $\forall v:~w^0_v\leftarrow \Delta_\mathrm{v}(y_v)$, $\forall c:~z^1_c\leftarrow \Delta_\mathrm{c}(s^0_c)$, $l\leftarrow 1$
\WHILE{$\mathbf{s}^l\neq \mathbf{0}$ and $l<l^{\mathrm{m}}_\mathcal{F}$}
\STATE $\forall v:~w_v^l\leftarrow f(w_v^{l-1},\chi_{0_{\mathrm{p}}}^{l}(v), \chi_{0_{\mathrm{n}}}^{l}(v), \chi_{1_{\mathrm{p}}}^{l}(v),\chi_{1_{\mathrm{n}}}^{l}(v))$;
\STATE $\forall c:~z_c^{l+1}\leftarrow \Phi(s_c^{l-1},s^{l}_c)$;
\STATE $l\leftarrow l+1$;
\ENDWHILE
\end{algorithmic}
\end{algorithm}
\end{defn}

What makes a TBF algorithm novel is that a variable node has ``strength'' and a check node's reliability is evaluated based on its state in the previous iteration.
\section{Motivation}\label{sect_mot}
Consider a collection $\mathscr{A}$ of iterative decoding algorithms for LDPC codes. Let us assume for a moment that the set of all uncorrectable error patterns for each and every algorithm in $\mathscr{A}$ is known. More precisely, in the context of LDPC codes, we assume that the induced subgraphs on such error patterns can be enumerated for each decoding algorithm. This naturally suggests the use of a decoder $\mathcal{D}$ which employs multiple algorithms drawn from $\mathscr{A}$. The basis for this use of multiple algorithms is rather simple: If different algorithms are capable of correcting different error patterns, then a decoder employing a set of properly selected algorithms can achieve provably better error performance than any single-algorithm decoder.
Disappointingly, the above hypothetical assumption is not valid for most iterative algorithms. For message passing algorithms such as the SPA, there is no simple criterion to verify weather or not an arbitrary error pattern is correctable, much less an explicit methodology to design a decoder which employs multiple algorithms in a collaborative manner.

Interestingly, for TBF algorithms, we are able to establish a framework to analyze and enumerate all uncorrectable error patterns, and this is the main contribution of this paper. In particular, we characterize the decoding failures of TBF algorithms by redefining trapping sets and introducing the definition of trapping set profiles. It is an important property of the newly defined trapping sets that enable us to enumerate them using a recursive procedure. We remark that the enumeration of trapping sets is code independent. More importantly, the concept and explicit construction of trapping set profiles allow rigorous selections of multiple algorithms which can collectively correct a fixed number of errors with high probability. Given that the selection of multiple algorithms would become straightforward once the trapping sets/trapping set profiles have been defined and constructed, we devote a considerable portion of the paper to introducing these two objects. We also focus on giving criteria for selecting algorithms rather than explicitly describing the selection process.
\section{Trapping Sets and Trapping Set Profiles}\label{sect_analysis}
\subsection{Trapping Sets of TBF Algorithms}
Although the term trapping set was originally defined as a set of variable nodes that are not eventually correctable by an iterative decoding algorithm \cite{errorFloor_richarson}, in the literature it has been used more frequently to refer to a \textit{combinatorially defined subgraph} that \textit{may} be harmful to decoding. The justification for this less rigorous use of terminology is that the variable node set of a so-called trapping set (a subgraph) would be an actual set of non-eventually-correctable variable nodes if the parallel bit flipping algorithm were used (see \cite{NCMV_11_IT} for details). Examples of
such trapping sets are fixed sets \cite{NCMV_11_IT} and absorbing sets \cite{absorbingSet_dolecek}. For TBF algorithms, failure analysis can no longer solely rely on these combinatorial objects. For certain TBF algorithms, the smallest subgraphs that cause decoding failures are neither absorbing sets nor fixed sets. We therefore (re)define the notion of a trapping set for TBF algorithms, as we now explain. We first introduce the following definition on failures of a TBF algorithm.
\begin{defn}
Consider a TBF algorithm $\mathcal{F}$ and a Tanner graph $G$. Let $V_\mathrm{e}$ denote the set of variable nodes that are initially corrupt and let $I$ denote the induced subgraph on $V_\mathrm{e}$. If the algorithm $\mathcal{F}$ does not converge on $G$ after $l^{\mathrm{m}}_\mathcal{F}$ iterations, then we say that \textit{$\mathcal{F}$ fails on the subgraph $I$ of $G$}.
\end{defn}

It can be seen that the decoding failure of $\mathcal{F}$ is defined with the knowledge of the induced subgraph on the set of initially corrupt variable nodes. To characterize failures of $\mathcal{F}$, a collection of all induced subgraphs $I$ must be enumerated. While this is difficult in general, for practically important cases of small numbers of initial errors (less than 8) and small column-weight codes ($d_\mathrm{v}= 3$ or 4), the enumeration of such induced subgraphs is tractable.

Consider a given Tanner graph $I$. Let $\mathscr{E}_I(\mathcal{F})$ denote a set of Tanner graphs containing a subgraph $J$ isomorphic to $I$ such that $\mathcal{F}$ fails on $J$. Since $\mathscr{E}_I(\mathcal{F})$ is undeniably too general to be useful, we focus our attention on a subset $\mathscr{E}_I^{\mathrm{r}}(\mathcal{F})$ of $\mathscr{E}_I(\mathcal{F})$, described as follows. 

\begin{defn}\label{defn_Er}
Consider a Tanner graph $S_1\in \mathscr{E}_I(\mathcal{F})$ such that $\mathcal{F}$ fails on the subgraph $J_1$ of $S_1$. Then, $S_1\in \mathscr{E}_I^{\mathrm{r}}(\mathcal{F})$ if there \textit{does not} exist $S_2\in \mathscr{E}_I(\mathcal{F})$ such that:
\begin{enumerate}
\item $\mathcal{F}$ fails on the subgraph $J_2$ of $S_2$, and
\item there is an isomorphism between $S_2$ and a proper subgraph of $S_1$ under which the variable node set $V(J_2)$ is mapped into the variable node set $V(J_1)$.
\end{enumerate}
\end{defn}



Now we are ready to define trapping sets and trapping set profiles of a TBF algorithm.

\begin{defn}
If $S\in \mathscr{E}_I^{\mathrm{r}}(\mathcal{F})$ then $S$ is a trapping set of $\mathcal{F}$. $I$ is called an inducing set of $S$. $\mathscr{E}_I^{\mathrm{r}}(\mathcal{F})$ is called the trapping set profile with inducing set $I$ of $\mathcal{F}$.
\end{defn}

The following proposition states an important property of a trapping set.

\begin{prop}\label{allCProp}
Let $S$ be a trapping set of $\mathcal{F}$ with inducing set $I$. Then, there exists at least one induced subgraph $J$ of $S$ which satisfies the following properties: 
\begin{enumerate}
\item $J$ is isomorphic to $I$, and
\item $\mathcal{F}$ fails on $J$ of $S$, and
\item Consider the decoding of $\mathcal{F}$ on $S$ with $V(J)$ being the set of initially corrupt variable nodes. Then, for any variable node $v\in V(S)$, there exist an integer $0\leq l \leq l^{\mathrm{m}}_\mathcal{F}$ such that $w^l_v\in\{1_\mathrm{s},1_\mathrm{w}\}$.
\end{enumerate}
\end{prop}
\IEEEproof
The proof is omitted due to page limits.
\endIEEEproof

From Proposition \ref{allCProp}, one can see that the trapping set profile $\mathscr{E}_I^{\mathrm{r}}(\mathcal{F})$ of $\mathcal{F}$ contains the graphs that are most ``compact.'' We consider these graphs most compact because for at least one $J$ isomorphic to $I$, the decoding of $\mathcal{F}$ on such a graph with $V(J)$ being the set of initially corrupt variable nodes could be made successful \textit{by removing any variable node of the graph}. This special property of trapping sets is the basis for an explicit recursive procedure to obtain all trapping sets up to a certain size, which compensates for the lack of a fully combinatorial characterization of trapping sets. We remark that for certain reasonably good algorithms, the necessary condition for a Tanner graph to be a trapping set can be easily derived. Before describing the recursive procedure for constructing trapping set profiles, we state the following proposition, which gives a sufficient condition for the convergence of an algorithm $\mathcal{F}$ on a Tanner graph $G$.
\begin{prop}\label{profF}
Consider decoding with an algorithm $\mathcal{F}$ on a Tanner graph $G$. Let $V_{\mathrm{e}}$ be the set of initially corrupt variable nodes and $I$ be the induced subgraph on $V_{\mathrm{e}}$. Then, algorithm $\mathcal{F}$ will converge after at most $l^{\mathrm{m}}_\mathcal{F}$ decoding iterations if there does not exist a subset $V_{\mathrm{s}}$ of $V(G)$ such that $V_{\mathrm{s}} \supset V_{\mathrm{e}}$ and the induced subgraph on $V_{\mathrm{s}}$ is isomorphic to a graph in $\mathscr{E}_I^{\mathrm{r}}(\mathcal{F})$.
\end{prop}
\IEEEproof Follows from the definition of $\mathscr{E}_I^{\mathrm{r}}(\mathcal{F})$.
\endIEEEproof
\textit{Remark:} Proposition \ref{profF} only gives a sufficient condition because the existence of $V_{\mathrm{s}}\subset V(G)$ which satisfies the above-mentioned conditions does not necessarily indicate that $G\in\mathscr{E}_I(\mathcal{F})$.
\subsection{Constructing a Trapping Set Profile}\label{sect_algo}
The  recursive procedure for constructing a trapping set profile $\mathscr{E}_I^{\mathrm{r}}(\mathcal{F})$ relies on Proposition \ref{allCProp}. Let us assume that we are only interested in trapping sets with at most $n^\mathrm{max}$ variable nodes. Consider the decoding of $\mathcal{F}$ on a Tanner graph $I$ with $V(I)$ being the set of initially corrupt variable nodes. Let $n_I = |V(I)|$. If $\mathcal{F}$ fails on the subgraph $I$ of $I$ then $\mathscr{E}_I^{\mathrm{r}}(\mathcal{F})=\{I\}$ and we have found the trapping set profile. If $\mathcal{F}$ does not fail on the subgraph $I$ of $I$, then we expand $I$ by recursively adding variable nodes to $I$ until a trapping set is found. During this process, we only add variable nodes that become corrupt at the end of a certain iteration. 

Consider all possible bipartite graphs obtained by adding one variable node, namely $v_{n_I+1}$, to the graph $I$ such that when the decoding is performed on these graphs with $V(I)$ being the set of initially corrupt variable nodes, the newly added variable node is a corrupt variable node at the end of the \textit{first iteration}, i.e., $w^1_{v_{n_I+1}}\in\{1_{\mathrm{w}},1_{\mathrm{s}}\}$. Let $\mathbf{O}_I$ denote the set of such graphs. Take one graph in $\mathbf{O}_I$ and denote it by $U$. Then, there can be two different scenarios in this step. First, $\mathcal{F}$ does not fail on the subgraph $I$ of $U$. In this case, $U$ is certainly not a trapping set and we put $U$ in a set of Tanner graphs denoted by $\mathbf{E}^1_I$. Second, $\mathcal{F}$ fails on the subgraph $I$ of $U$. In this case, $U$ can be a trapping set and a test is carried out to determine if $U$ is indeed one. If $U$ is not a trapping set then it is discarded. We complete the formation of $\mathbf{E}^1_I$ by repeating the above step for all other graphs in $\mathbf{O}_I$. 

Let us now consider a graph $U\in\mathbf{E}^1_I$. Again, we denote by $\mathbf{O}_U$ the set of Tanner graphs obtained by adding one variable node, namely $v_{n_I+2}$, to the graph $U$ such that when the decoding is performed on these graphs with $V(I)$ being the set of initially corrupt variable nodes, the newly added variable node is a corrupt variable node at the end of the first iteration, i.e., $w^1_{v_{n_I+2}}\in\{1_{\mathrm{w}},1_{\mathrm{s}}\}$. It is important to note that the addition of variable node $v_{n_I+2}$, which is initially correct, cannot change the fact that variable node $v_{n_I+1}$ is also corrupt at the end of the first iteration. This is because the addition of correct variable nodes to a graph does not change the states of the existing check nodes and the decoding dynamic until the newly added variable nodes get corrupted. Similar to what have been discussed before, we now take a graph in $\mathbf{O}_U$ and determine if it is a trapping set, or it is to be discarded, or it is a member of the set of Tanner graph $\mathbf{E}^2_I$. By repeating this step for all other graphs in $\mathbf{E}^1_I$, all graphs in $\mathbf{E}^2_I$ can be enumerated. In a similar fashion, we obtain $\mathbf{E}^3_I, \mathbf{E}^4_I,\ldots,\mathbf{E}^{(n^\mathrm{max}-n_I)}_I$. For the sake of convenience, we also let $\mathbf{E}^0_I = \{I\}$.

At this stage, we have considered one decoding iteration on $I$. It can be seen that if $S$ is a trapping set with at most $n^\mathrm{max}$ variable nodes then either $S$ has been found, or $S$ must contain a graph in $\bigcup_{i=0}^{(n^{\mathrm{max}}-n_I-1)}{\mathbf{E}^{i}_I}$. Therefore, we proceed by expanding graphs in $\mathbf{E}_I = \bigcup_{i=0}^{(n^{\mathrm{max}}-n_I-1)}{\mathbf{E}^{i}_I}$.

Let $K$ denote a Tanner graph in $\mathbf{E}_I = \bigcup_{i=0}^{(n^{\mathrm{max}}-n_I-1)}{\mathbf{E}^{i}_I}$. We now repeat the above graph expanding process with $K$ being the input. Specifically, we first obtain $\mathbf{O}_K$, which is defined as the set of all Tanner graphs obtained by adding one variable node $v_{n_K+1}$ to the graph $K$ such that when decoding is performed on these graphs with $V(I)$ being the set of initially corrupt variable nodes, the newly added variable node is a corrupt variable node at the end of the \textit{second iteration}, but not a corrupt variable node at the end of the first iteration, i.e., $w^1_{v_{n_K+1}}\in\{0_{\mathrm{w}},0_{\mathrm{s}}\}$ and $w^2_{v_{n_K+1}}\in\{1_{\mathrm{w}},1_{\mathrm{s}}\}$. Graphs in $\mathbf{O}_K$ that are not trapping sets are either discarded or to form the set $\mathbf{E}^1_K$. By recursively adding variable nodes, graphs in $\mathbf{E}^2_K, \mathbf{E}^3_K,\ldots,\mathbf{E}^{n^\mathrm{max}-n_I}_K$ are enumerated.

One can see that there are two recursive algorithms. The first algorithm enumerates graphs in $\mathbf{E}_K = \bigcup_{i=0}^{(n^{\mathrm{max}}-n_I)}{\mathbf{E}^{i}_K}$ for a given graph $K$ by recursively adding variable nodes. The second algorithm recursively calls the first algorithm to enumerates graphs in $\mathbf{E}_K = \bigcup_{i=0}^{(n^{\mathrm{max}}-n_I)}{\mathbf{E}^{i}_K}$ for each graph $K$ in $\mathbf{E}_I = \bigcup_{i=0}^{(n^{\mathrm{max}}-n_I-1)}{\mathbf{E}^{i}_I}$. Each recursion of the second algorithm corresponds to a decoding algorithm. As a result, the trapping set profile is obtained after $l^\mathrm{m}_\mathcal{F}$ recursions of the second algorithm.
\section{Selecting TBF Algorithms}\label{sect_select}
Due to page limits, we only summarize the most important criteria for selecting TBF algorithms. Let us first briefly discuss the number of possible algorithms.
\subsection{On the Number of Algorithms}
Let $\mathcal{Q}$ be the set of all functions from $\mathcal{A}_v\times \Xi_{d_\mathrm{v}}\rightarrow \mathcal{A}_v$ that satisfy the symmetry and the irreducibility condition. Due the symmetry condition, $|\mathcal{Q}|\leq 4^{2\times|\Xi_{d_\mathrm{v}}|}$. There are two possible values of $\Delta_\mathrm{v}$, and two possible values of $\Delta_\mathrm{c}$. However, with a given $\Delta_\mathrm{c}$, the two sets of algorithms $\mathcal{F}$ that correspond to two possible $\Delta_\mathrm{v}$ are identical (as $0_{\mathrm{s}}$ and $0_{\mathrm{w}}$, $1_{\mathrm{s}}$ and $1_{\mathrm{w}}$ can be interchanged). Consequently, if we disregard the maximum number of iterations, then $|\mathscr{F}|=2|\mathcal{Q}|\leq 2^{(4|\Xi_{d_\mathrm{v}}|+1)}$. One can easily show that $|\Xi_{d_\mathrm{v}}| =  \binom{d_\mathrm{v}+3}{3}$. Therefore, an upper-bound on the number of TBF algorithms is:
\begin{eqnarray}
|\mathscr{F}|\leq 2^\frac{2{d_\mathrm{v}}^3+12{d_\mathrm{v}}^2+22{d_\mathrm{v}}+15}{3}.\label{eqn_up}
\end{eqnarray}
For example, this upper-bound is $2^{81}$ when ${d_\mathrm{v}}=3$, and is $2^{141}$ when ${d_\mathrm{v}}=4$.%

Due to the huge number of possible algorithms, it is necessary to focus on a small subset of algorithms. This subset of algorithms may be obtained by imposing certain constraints on the function $f$. One example of such a constraint is as follows: if $f(0_\mathrm{s},\xib)\in \{1_\mathrm{w},1_\mathrm{s}\}$ then $f(0_\mathrm{w},\xib)\in \{1_\mathrm{w},1_\mathrm{s}\}$. This constraint requires that when a strong zero variable node is flipped with a given combination of check nodes, a weak variable node is also flipped with the same check node combination. Other constraints on $f$ are derived by analyzing possible transitions of variable nodes and check nodes for a small number of iterations.
\subsection{Selecting a TBF Algorithm}
We first discuss the main criterion to select one algorithm among all possible algorithms. Let $n^\mathrm{min}_{I,\mathcal{F}}$ be the smallest number of variable nodes of Tanner graphs in $\mathscr{E}_I^{\mathrm{r}}(\mathcal{F})$. We would like to select an algorithm $\mathcal{F}$ such that $n^\mathrm{min}_{I,\mathcal{F}}$ is maximized. The justification for this selection criterion relies on the following proposition, whose proof is omitted due to page limits.
\begin{prop}\label{profG}
Given three random Tanner graph $G,S_1,S_2$ with $0<|V(S_1)|<|V(S_2)|<|V(G)|$, the probability that $G$ contains $S_2$ is less than the probability that $G$ contains $S_1$.
\end{prop}

From Proposition \ref{profG}, one can see that the larger the number $|V(S)|$ of a given Tanner graph $S$ is, the easier it would be (if at all possible) to construct a Tanner graph $G$ that does not contain $S$. Therefore, a larger $n^\mathrm{min}_{I,\mathcal{F}}$ means that the sufficient condition for the convergence of $\mathcal{F}$ can be met with higher probability. In this sense, an algorithm $\mathcal{F}$ with a larger $n^\mathrm{min}_{I,\mathcal{F}}$ is more favorable.

If for two algorithms $\mathcal{F}_1$ and $\mathcal{F}_2$, $n^\mathrm{min}_{I,\mathcal{F}_1}$ = $n^\mathrm{min}_{I,\mathcal{F}_2}$, then one can derive other comparison criteria based on $\mathscr{E}_I^{\mathrm{r}}(\mathcal{F}_1)$ and $\mathscr{E}_I^{\mathrm{r}}(\mathcal{F}_2)$, and/or compare $\mathcal{F}_1$ and $\mathcal{F}_2$ with a different assumption of $I$. For example, the probability of a graph $G$ containing a trapping set $S$ can be also be evaluated based on $|C(S)|$.
\subsection{Selecting Multiple TBF Algorithms}
We now consider the problem of selecting of multiple algorithms. The basis for this selection is that one should select good individual algorithms with diverse trapping set profiles. In this paper, we only consider decoder $\mathcal{D}$ with algorithms $\mathcal{F}_1,\mathcal{F}_2,\ldots,\mathcal{F}_p$ operating in parallel, i.e., the received vector of the channel is the input vector for all algorithms. Note that one can also use trapping set profiles to select algorithms that operate in serial, i.e., the output from one algorithm is the input to another. For a decoder $\mathcal{D}$ that employs parallel algorithms, the concept of trapping sets and trapping set profiles can be defined in the same manner as trapping sets and trapping set profiles for a single TBF algorithm. One can easily modify the recursive procedures given in Section \ref{sect_algo} to generate trapping set profiles of the decoder $\mathcal{D}$. Then, $\mathcal{D}$ can be designed with the same criterion discussed in the previous subsection.

\textit{Remark:} Knowledge on the Tanner graph of a code $\mathcal{C}$ can be used in the selection of algorithms. For example, if it is known that the Tanner graph of $\mathcal{C}$ does not contain a certain subgraph $Y$, then all graphs containing $Y$ must be removed from a trapping set profile.
\section{Numerical Results}\label{sect_sim}
As an example, we describe a selection of TBF algorithms for regular column-weight-three LDPC codes with girth $g=8$. For simplicity, we let $\Delta_\mathrm{v} = (0_\mathrm{s},1_\mathrm{s})$,  $\Delta_\mathrm{c} = (0_\mathrm{p},1_\mathrm{p})$ and $l^\mathrm{m}_\mathcal{F} = 30$ for all algorithms. By imposing certain constraints on the functions $f$, we obtain a set of $21,962,496$ TBF algorithms. Out of these, there are $360,162$ algorithms which can correct any weight-three error pattern. Such an algorithm is capable of correcting any weight-three error pattern because its trapping set profile $\mathscr{E}_I^{\mathrm{r}}(\mathcal{F})$ with any inducing set $I$ containing three variable nodes is empty. Since all weight-three error patterns can be corrected with a single algorithm, our next step is to select a collection of algorithms which can collectively correct weight-four and -five error patterns with high probability. To achieve this goal, we construct all trapping set profiles with inducing sets containing four and five variable nodes for each algorithm. Note that there are 10 possible inducing sets (Tanner graphs with girth $g=8$) containing four variable nodes and 24 possible inducing sets containing five variable nodes. Hence, for each algorithm, we construct a total of 34 trapping set profiles. From the trapping set profiles of all algorithms, we select a collection of 35 algorithms based on the criterion mentioned in the previous section. Then, we simulate the performance of a decoder $\mathcal{D}$ which employs these algorithms in parallel. The maximum total number of iterations of $\mathcal{D}$ is $35\times 30 = 1050$. 

\begin{figure}
\centering
\includegraphics[height = 2.85in]{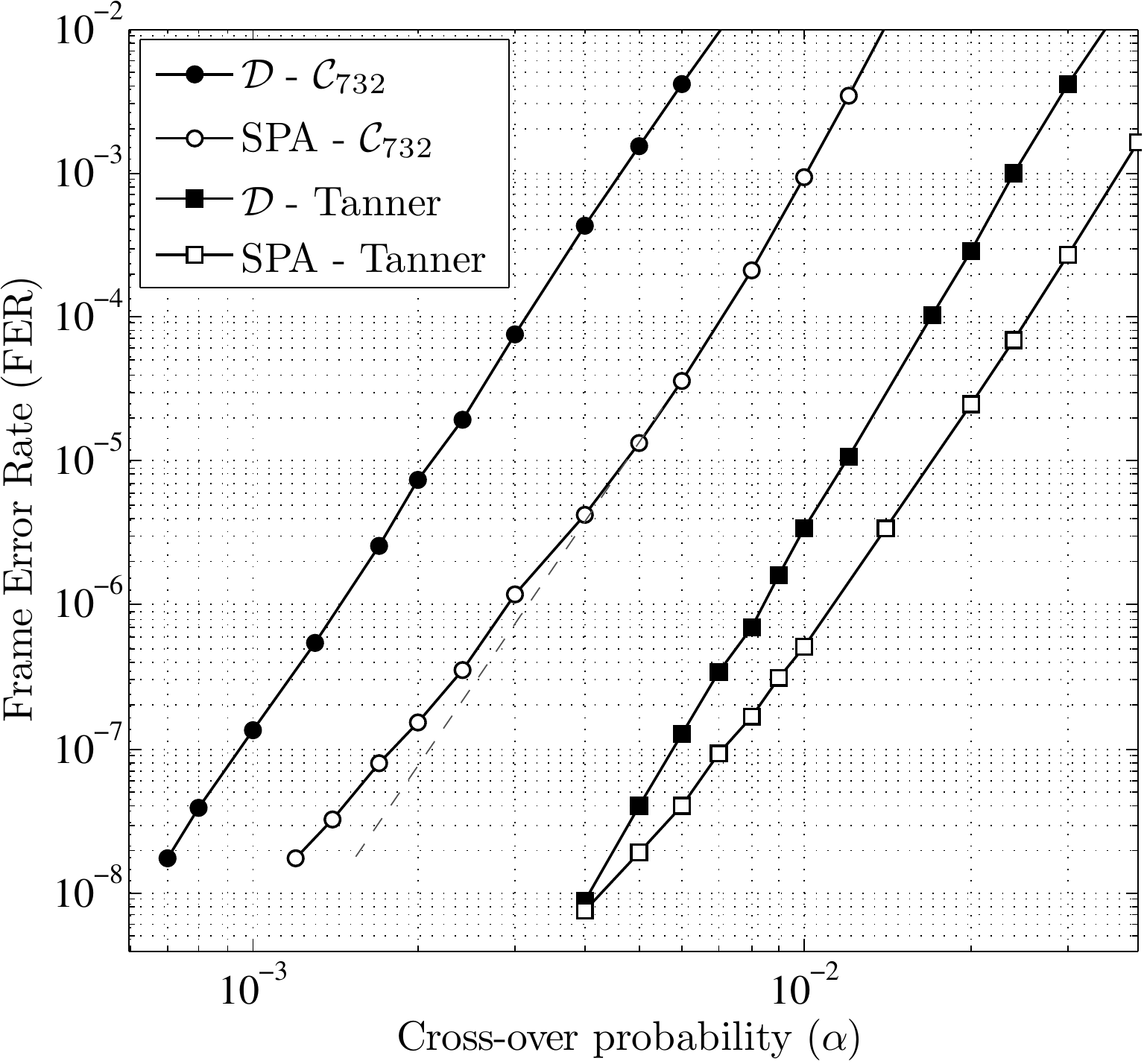}
\caption{Frame error rate performance of the decoder $\mathcal{D}$.}
\label{fig_fer1}
\vspace{-0.05in}
\end{figure}

Figure \ref{fig_fer1} shows the frame error rate (FER) performance of $\mathcal{D}$ on the $(155,64)$ Tanner code. This code has $d_\mathrm{v} = 3$, $d_\mathrm{c} = 5$ and minimum distance $d_\mathrm{min} = 20$. For comparison, the FER performance of the SPA with a maximum of 100 iterations is also included. It can be seen that the FER performance of $\mathcal{D}$ approach (and might surpasses) that of the SPA in the error floor region. It is also important to note that if we eliminate all trapping sets containing subgraphs that are not present in the Tanner graph of this code, then all the obtained trapping set profiles are empty. This indicates that $\mathcal{D}$ can correct any error pattern up to weight 5 in the Tanner code.

Figure \ref{fig_fer1} also shows the FER performance of $\mathcal{D}$ on a quasi-cyclic code $\mathcal{C}_{732}$ of length $n=732$, rate $R=0.75$ and minimum distance $d_\mathrm{min} = 12$. The FER performance of the SPA is also included for comparison. It can be seen that the slope of the FER curve of $\mathcal{D}$ in the error floor region is higher than that of the SPA. Finally, we remark that the slope of the FER curve of $\mathcal{D}$ in the error floor region is between 5 and 6, which indicates that $\mathcal{D}$ can correct error patterns of weight 4 and 5 with high probability. This also agrees with the fact that in our simulation, no weight-four error pattern that leads to decoding failure of $\mathcal{D}$ was observed.

We remark that the implementation of TBF algorithms operating in parallel can be done with a relatively small number of common logic gates. For example, if a decoder $\mathcal{D}$ employs both the TBFA1 and the TBFA2 given in \cite{NMV_11_ISIT}, then the implementation of the variable node updates require less than 800 AND-gate inputs and 100 OR-gate inputs. In comparison, the implementation of a 6-bit adder requires 2196 AND-gate inputs and 355 OR-gate inputs while that of a 6-bit comparator requires 1536 AND-gate inputs and 190 OR-gate inputs. One can also expect that the complexity introduced by an additional algorithm would decrease as the number of algorithms increases, because many min-terms in the variable node update logic functions would be already available. More details will be provided in the journal version of this paper. 

\section*{Acknowledgment}
This work is funded by NSF under the grants CCF-0963726, CCF-0830245.
\bibliographystyle{IEEEtran}

\end{document}